\documentclass[aps,prd,twocolumn,amsmath,superscriptaddress,amssymb,showpacs,floatfix,nofootinbib,longbibliography,preprintnumbers,longbibliography]{revtex4-1}
\pdfoutput=1
\usepackage{graphicx}
\usepackage{bm}
\usepackage{times}
\usepackage{slashed}
\usepackage{color}
\usepackage{aas_macros} 
\usepackage{slashed}
\usepackage{lipsum}
\usepackage{subfigure}
\usepackage{multirow} 
\usepackage{amsmath}
\usepackage{array} 
\usepackage{varwidth} 

\usepackage{hyperref}

\hypersetup{
     colorlinks   = true,
     citecolor    = magenta,
     urlcolor     = magenta,
     linkcolor    = magenta
}

\newcommand{\be}{\begin{equation}}
\newcommand{\ee}{\end{equation}}
\newcommand{\bea}{\begin{eqnarray}}
\newcommand{\eea}{\end{eqnarray}}

\begin{document}

\title{Constraints on Ultra-heavy DM from TeV-PeV gamma-ray diffuse measurements}

\author{Manuel Rocamora}
\thanks{{\scriptsize Email}: \href{mailto:manuel.rocamora@uibk.ac.at}{manuel.rocamora@uibk.ac.at};}
\affiliation{Universität Innsbruck, Institut für Astro- und Teilchenphysik, Technikerstr. 25/8, 6020 Innsbruck, Austria}
\author{Pedro De La Torre Luque}
\thanks{{\scriptsize Email}: 
\href{mailto:pedro.delatorre@uam.es}{pedro.delatorre@uam.es}, {\scriptsize ORCID}: \href{ http://orcid.org/0000-0002-4150-2539}{0000-0002-4150-2539}}
\author{Miguel A. Sánchez-Conde}
\thanks{{\scriptsize Email}: 
\href{mailto:miguel.sanchezconde@uam.es}{miguel.sanchezconde@uam.es}}
\affiliation{Departamento de F\'{i}sica Te\'{o}rica, M-15, Universidad Aut\'{o}noma de Madrid, E-28049 Madrid, Spain}
\affiliation{Instituto de F\'{i}sica Te\'{o}rica UAM-CSIC, Universidad Aut\'{o}noma de Madrid, C/ Nicol\'{a}s Cabrera, 13-15, 28049 Madrid, Spain}

\begin{abstract}
Recent experiments have measured the Galactic $\gamma$-ray diffuse emission up to PeV energies, opening a window to study acceleration of Galactic cosmic rays and their propagation up to the cosmic-ray knee. Furthermore, these observations provide a powerful tool to set strong constraints into very-heavy dark matter particles, with masses in the TeV-PeV range. In this paper, we explore the potential of the newest observations of diffuse emissions at the Galactic plane from HAWC and LHAASO to probe this kind of dark matter over a wide mass range. 
Here, we model secondary emissions (inverse-Compton) from the electrons and positrons produced in the annihilation/decay of dark matter, on top of their prompt $\gamma$-ray emission, including the effects of absorption of high-energy photons via pair production. Furthermore, we show that including the astrophysical backgrounds (namely diffuse emission from cosmic-ray collisions or emission from unresolved sources) can significantly improve these limits.
We find that the new measurements provided, specially by LHAASO with the combination of the WCDA and KM2A detectors, allow us to set strong constraints in decaying dark matter, being competitive and even improving the strongest constraints at the moment. 
We also highlight that these regions lead to constraints that are less affected by uncertainties from the dark matter distribution and discuss how CTA north and SWGO will be able to improve limits in this mass range.

\end{abstract}

\maketitle

\section{Introduction}

The nature of the Dark Matter (DM) remains one of the most intriguing mysteries still today. Despite all the efforts dedicated in recent years to search for signatures of DM (either directly or indirectly), no clear counterpart to its gravitational evidences has been found so far~\cite{cirelli2024darkmatter, Nesti_2023}. 
A variety of well-motivated models for DM as a particle has been proposed, being the WIMP (Weakly Interactive Massive Particles) the paradigm that has received more attention, unfortunately without success~\cite{Steigman_2012, Miener:2022zws, justino2024constraintsinertdoubletmodel, Zuriaga_Puig_2023, DelaTorreLuque:2024ozf, PhysRevD.109.L041301, FermiLAT:2011ozd}. However, recent experimental developments and improvements in detection techniques continue to reduce the available parameter space of the particle candidates that can be compatible with DM. In particular, the $\gamma$-ray community has played a fundamental role in the indirect search of DM signals. Remarkably, recent $\gamma$-ray experiments are providing measurements that reach PeV energies~\cite{Cao_2023}, which allows us to probe some of the heaviest DM candidates that are out of the reach of the Large Hadron Collider or even current direct detection experiments. 

Indeed, although annihilation of thermal WIMPs is expected to be only possible for DM masses up to $\sim100$~TeV, due to the unitarity bound~\cite{Unitarity, Smirnov_2019}, decay is allowed for any mass. Furthermore, besides WIMPs, which still remain as the most compelling DM candidate, other very-heavy DM (VHDM) candidates, with masses in the TeV-PeV range, can be produced from fluctuations of the gravitational fields or phase transitions in the early Universe, and thermalize via freeze-out or freeze-in~\cite{Carney_2023, Benakli_1999, Hambye_2020, Chianese_2017, Higaki_2014, Contino_2019, Kim_2019, Chung_1998, kolb1998wimpzillas, Kolb_2017, CompositeDM, Faraggi_2002, Boddy_2014, Carenza_2023, Cembranos_2003}. 
Annihilation and decay of VHDM will produce very-high-energy (VHE) $\gamma$ rays that suffer substantial absorption in their propagation due to pair-production with Cosmic Microwave Background (CMB) light. Therefore, at these energies, the $\gamma$-ray flux becomes significantly attenuated far from the source and the horizon of observation is below $100$~kpc~\cite{Lipari_Vernetto}. This is the reason why extragalactic $\gamma$-ray observations become ineffective for conducting indirect DM searches at masses above the TeV.

The current generation of ground-based $\gamma$-ray experiments has been able to probe DM annihilation and decay up to DM masses of $\sim 100$~TeV~\cite{ACCIARI201838, galaxies8010025}. The Cherenkov Telescope Array Observatory (CTAO) is expected to measure $\gamma$-ray fluxes up to around $300$~TeV and provide a factor of $10$ better sensitivity than the current generation of ground-based TeV experiments~\cite{2019_CTA_Book}. Similar capabilities are also expected from the Southern Wide-field Gamma-ray Observatory (SWGO)~\cite{SWGO}, which will offer wide-field coverage of a significant portion of the southern sky. The Fermi Large Area Telescope (Fermi-LAT) on board the NASA Fermi satellite has provided strong constraints on VHDM from the study of the low-energy photons~\cite{Ackermann_2012} produced directly in the annihilation/decay or from the (secondary) inverse-Compton (IC) emission associated to the e$^+$e$^-$ products~\cite{Li_2016}.

The High-Altitude Water Cherenkov Observatory (HAWC)~\cite{HAWC:2016bfh} and the Large High Altitude Air Shower Observatory (LHAASO) collaborations~\cite{LHAASO:2019qtb} have recently released a new set of ultra-high energy measurements in different Galactic regions. Although the precision of this data is still limited due to poor statistics, it already opens up the possibility to push further the constraints on the lifetime or annihilation rate of DM in the TeV-PeV mass range. This is expected to be especially relevant for decaying DM models, given the mid longitude range of the datasets.

The goal of this work is to make use of these new datasets to constrain the possible existence of DM $\gamma$-ray signals and, in its absence, to expand the bounds to the highest DM masses explored. We will do so within a consistent framework in which we take full consideration of the prompt $\gamma$-ray emissions, secondary emissions (i.e. IC emission by secondary electrons and positrons) and the $\gamma$-ray absorption due to pair production on the CMB and Galactic radiation fields. First, we will obtain robust, conservative limits on the DM parameter space by making the minimal set of assumptions, and then, by modeling the Galactic diffuse astrophysical flux, we will be able to place even more stringent bounds.

The paper is organized as follows: first, we introduce the HAWC and LHAASO datasets we consider in Section~\ref{sec:Datasets}; then, we describe our modelling of the $\gamma$-ray emission and procedure to set the DM constraints in Section~\ref{sec:Method}. We discuss the compatibility of these $\gamma$-ray measurements with the signal expected from standard astrophysical mechanisms and how a dominant DM contribution seems incompatible with them, in Section~\ref{sec:results}, and set DM limits to finally conclude in Section~\ref{sec:discussion} with a highlight and discussion of the main findings.

\section{Galactic diffuse gamma-ray data sets at the TeV-PeV} 
\label{sec:Datasets}

We make use of the most recent $\gamma$-ray data in the TeV-PeV domain collected by HAWC and LHAASO.
HAWC is a VHE $\gamma$-ray observatory consisting in 300 Water-Cherenkov detectors, covering an area of 22,000 m$^2$ in Mexico. We make use of the latest published HAWC Galactic diffuse data from Ref.~\cite{HAWC_data}. The analysis covers an energy range from 300 hundred GeV to $100$ TeV of the Galactic plane region with $l \in [43^{\circ}, 73^{\circ}]$. The HAWC collaboration presented the average gamma-ray flux in the whole mentioned longitude range at $|b|<2^{\circ}$ (the inner plane) and $|b|<4^{\circ}$, as well as the flux of the three sub-regions, i.e. $l \in [43^{\circ}, 56^{\circ}]$, $[56^{\circ}, 64^{\circ}]$ and $[64^{\circ}, 73^{\circ}]$ . In addition to these averaged fluxes, they provided the longitudinal (in $3^{\circ}$ longitude bins at $|b|<2^{\circ}$ and $|b|<4^{\circ}$) and latitudinal emission profiles ($|b|<4.5^{\circ}$ in nine $1^{\circ}$ latitude bins, at $43^{\circ} < l < 73^{\circ}$). 

We also consider LHAASO data from the Water Cherenkov Detector Array (WCDA) and the Kilometer-2 Array (KM2A). LHAASO is a VHE $\gamma$-ray observatory, combining Water-Cherenkov detectors, charged particle detectors and muon detectors covering an area of 78,000 m$^2$ in China. In Ref.~\cite{Cao_2023}, the collaboration presented the averaged flux in the inner ($15^{\circ} < l < 125^{\circ}$) and outer ($125^{\circ} < l < 235^{\circ}$) Galactic plane ($|b| < 5^{\circ}$) from $10$~TeV to $1$~PeV, using a mask to remove the flux of the detected $\gamma$-ray sources. Later, they provided an update on this data in Ref.~\cite{WCDA_data} to expand the energy range down to 1 TeV.
As in the case of HAWC, they also provide latitude and longitude profiles for the emission in the analyzed regions, divided in $3-25$, $10$-$63$ and $63$-$1000$ TeV energy bands.

\begin{figure}[th!]
\includegraphics[width=0.95\columnwidth]{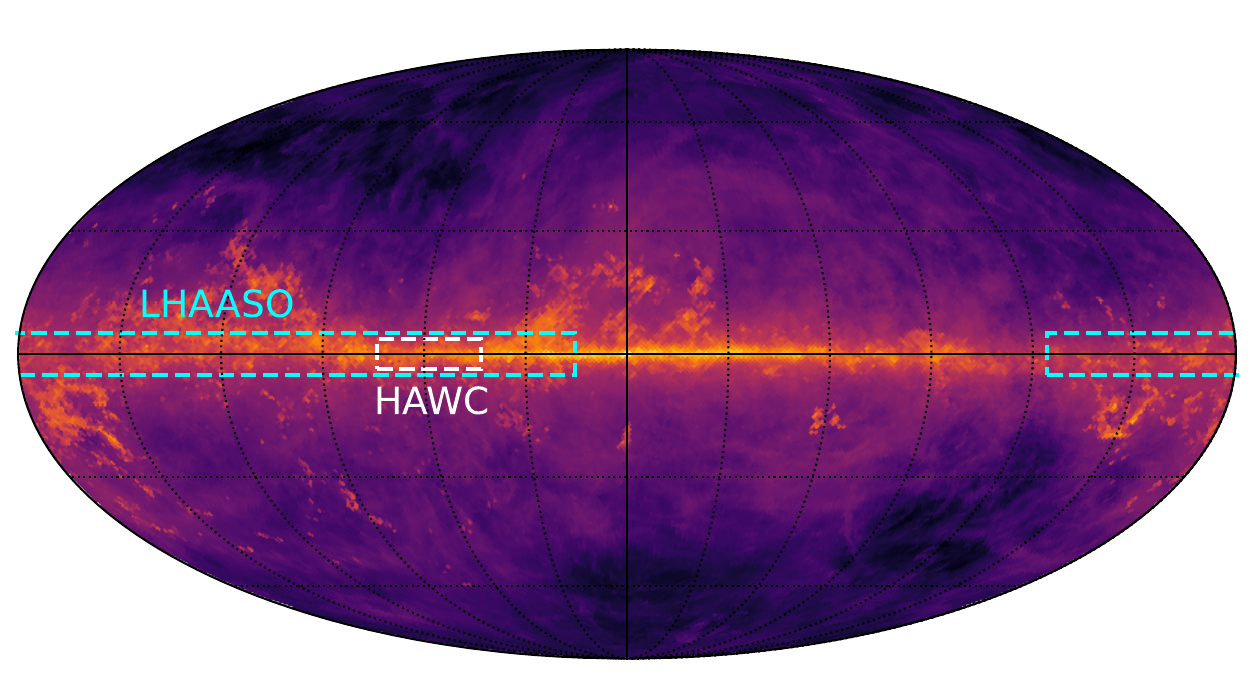}
\caption{Skymap depicting the two datasets considered in this work. The LHAASO Galactic plane region \cite{Cao_2023} is shown in blue, while the HAWC region \cite{HAWC_data} is in white. Shown in the background is the Galactic diffuse astrophysical model introduced in Section~\ref{sec:Method} and presented in Refs.~ \cite{DelaTorreLuque:2022ats,Luque_PeV}.}\vspace{-3mm}
\label{fig:Map}
\end{figure}

In Fig.~\ref{fig:Map} we display a map of the Galactic $\gamma$-ray diffuse emission at $10$~TeV, where we highlight the two regions covered by HAWC and LHAASO observations and used in our DM analysis. 
The combination of these datasets allows us to study a wide range of DM masses from a few TeV to tens of PeV. This requires both an accurate modelling of the prompt $\gamma$-ray from DM decay/annihilation and the secondary emission associated to the production of electrons and positrons by DM, as well as the absorption of high-energy photons. As explained below, these constraints can be significantly improved by means of the adoption of background models accounting for the Galactic diffuse astrophysical emission.

\section{Expected gamma-ray signals}
\label{sec:Method}

\textbf{DM induced direct production of $\gamma$-rays}:
The direct production of $\gamma$-rays from DM is determined by the DM distribution and the $\gamma$-ray emissivity. Regarding the distribution, throughout this paper we will use the Einasto DM density profile~\cite{einasto_prof}, which is described by
\begin{equation}
    \rho_\chi(r) = \rho_s ~ exp\left[-\frac{2}{\alpha_s}\left(\left(\frac{r}{r_s}\right)^{\alpha_s}-1\right)\right],
    \label{eq:einasto}
\end{equation}
where $\alpha_s$ = 0.17 , $r_s$ = 15.7 kpc~\cite{einasto_pars} and $\rho_s$ is determined with the local DM density, which we assume to be $\rho_{\odot}$ = 0.4 GeV/cm$^3$ \cite{Iocco_2011, Read_2014}. We adopt this DM density profile in order to allow for a more direct comparison with previous results, yet we explore the effect of using other DM density profiles in Appendix~\ref{sec:DM_prof}.

The gamma-ray emissivity depends on whether gamma rays are produced through DM annihilations or decays. If the production is via DM annihilations, it is described by:
\begin{equation}
    \epsilon_{ann}\,(E_{\gamma}, r) = \frac{\langle \sigma v \rangle_\chi m^2_\chi}{8\pi} \rho^2_{\chi}(r) \frac{dN}{dE_{\gamma}},
    \label{eq:ann}
\end{equation} 
where m$_\chi$ is the mass of the WIMP, $\langle \sigma v \rangle_\chi$ is the annihilation cross-section and $\rho_\chi$ is the DM density given by Eq.~(\ref{eq:einasto}). For decay, the emissivity is given by:
\begin{equation}
    \epsilon_{dec}\,(E_{\gamma}, r) = \frac{1}{4\pi m_{\chi} \tau_{\chi}}\rho_{\chi}(r) \frac{dN}{dE_\gamma},
    \label{eq:dec}
\end{equation}
where $\tau_{\chi}$ is the mean lifetime of the DM particle. In both Eqs.~(\ref{eq:ann}) and (\ref{eq:dec}), the gamma-ray spectrum generated per WIMP annihilation is dN/dE$_{\gamma}$, normalized such that its integral over energy is equal to 1. 
This injection spectrum is a crucial ingredient in these calculations. We use PPPC4DM~\cite{PPPC4_ID} as reference, which reaches DM masses of $100$~TeV for annihilation spectra\footnote{We note that the recently released CosmiXs package~\cite{Arina_2024} could lead to differences in the prompt DM signals of $\mathcal{O}(10\%)$ at most, and covers the same mass range, but we expect negligible changes in the constraints obtained}. For heavier masses, we make use of the {\tt HDMSpectra}\footnote{\url{https://github.com/nickrodd/HDMSpectra}} code~\cite{HDMSpectra}. We have checked that, for m$_{\chi} =100$~TeV, the differences in the DM $\gamma$-ray signal between using PPPC4DM and {\tt HDMSpectra} are negligible for the channels that we study in this work.

We note that the LHAASO mask used to remove source flux contributions and to measure the Galactic diffuse background emission removes a significant portion of the Galactic disk~\cite{Cao_2023}. Thus, we include it as well when comparing our predicted signals to LHAASO data. More precisely, after computing the $\gamma$-ray skymaps in the corresponding regions of interest, we mask out the circular regions as provided in the mask\footnote{\url{https://arxiv.org/src/2305.05372v2/anc/mask.txt}}.
This is not done for the analysis of HAWC data: in this case the effect of the mask is not so relevant given that the extension of the mask for each source is much smaller.

\textbf{Background astrophysical model:}
The Galactic diffuse emission at these energies is mainly composed by hadronic emission from $\pi^0$ decay (produced in the interactions of cosmic rays (CR), essentially H and He, with the interstellar gas), unresolved point-like and extended sources, and IC emission from high-energy diffuse electrons, which is expected to be subdominant in comparison to the others. In this work, we first derive conservative limits on diffuse DM signals by omitting any other source of diffuse emission in the Galaxy and, then, on top of the DM emission, we consider the astrophysical diffuse emission due to hadronic interactions as given by the ``Min'' model presented in Refs.~\cite{DelaTorreLuque:2022ats, Luque_PeV}, which is used as our benchmark background. This is a model built to fit the lower bound of the local CR measurements, set by the uncertainties among the different experiments close to the ``CR knee''. It also accounts for the hardening of the proton flux as a function of the Galactocentric radius, as measured by Fermi-LAT~\cite{Fermi-LAT:2012edv, Gaggero:2015xza, Fermi-LAT:2016zaq, Yang2016prd}, by using a spatial-dependent diffusion index.
This background estimation is not only conservative because of the use of the Min model, but because it does not include emission from unresolved sources. We note that a comparison of the morphology of the DM-induced $\gamma$-ray emission with that observed by LHAASO evidences that this contribution alone cannot reproduce the data for the latitude profile of the emission (see Fig.~\ref{fig:dm_profs} in Appendix \ref{sec:DM_prof}). Meanwhile, it is interesting to note that the longitude profile from decaying DM is very similar to that from the diffuse emission.

\begin{figure}[th!]
\includegraphics[width=0.95\columnwidth]{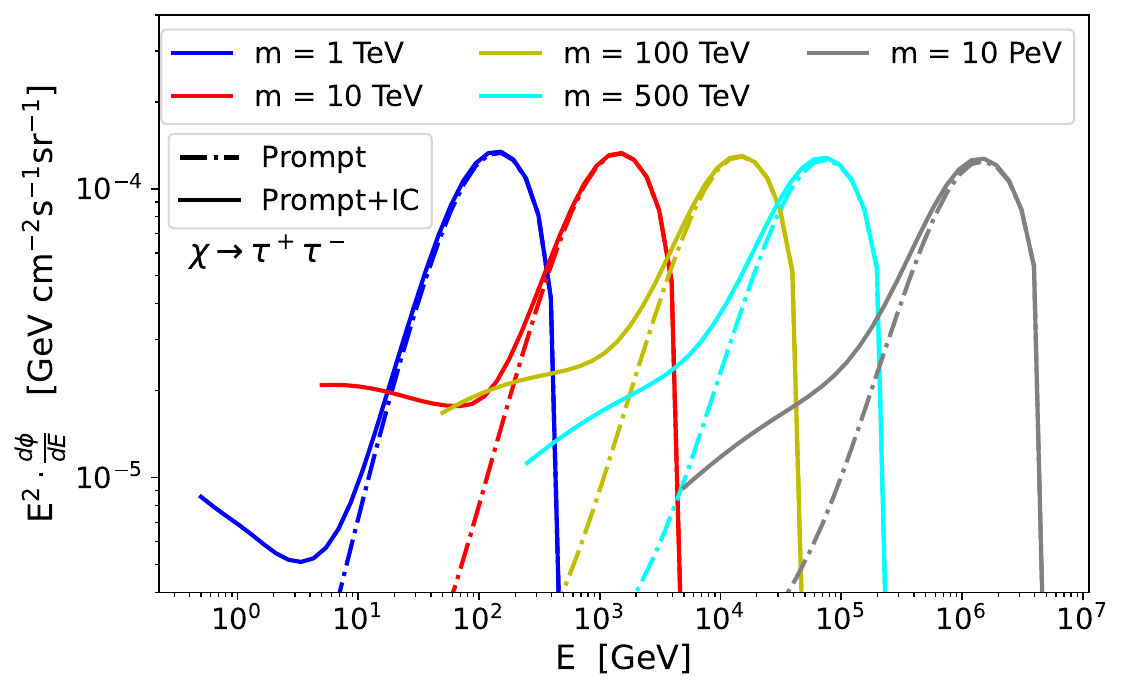}
\caption{Gamma-ray flux produced in DM decay into $\tau\tau$ via prompt emission (dash-dotted lines) and prompt plus secondary IC (solid) for a DM mass of 1 TeV (blue), 10 TeV (red), 100 TeV (yellow), 500 TeV (light blue) and 10 PeV (gray).} 
\label{fig:PromptIC}
\end{figure}

\textbf{Secondary emissions}: Together with the prompt production of $\gamma$-rays, DM annihilation or decay produce electrons and positrons whose interaction with the interstellar fields via IC produces a continuum $\gamma$-ray emission at energies below the peak of the prompt emission (see Fig.~\ref{fig:PromptIC}). A complete evaluation of the DM $\gamma$-ray signals requires accounting for the injection and propagation of these electrons-positrons to estimate such DM-induced secondary emissions. 
This contribution is more relevant in leptonic annihilation/decay channels, where their electron-positron yield is higher and these electrons/positrons carry more energy from the parent particle.
This has been used in the past to set constraints on heavy DM particles~\cite{Song_2024, leung2023improvinghawcdarkmatter}.
It is important to note that this secondary $\gamma$-ray emission peaks at lower energies than the prompt emission, becoming dominant at low energies because the prompt emission drops off rapidly below the. For the data sets considered in this work, this process becomes leading for DM masses around 2 orders of magnitude higher than the maximum energy of the measurements (see Fig.~\ref{fig:PromptIC}). 

Recently, Ref.~\cite{leung2023improving} derived DM constraints from HAWC flux upper-limits~\cite{Abeysekara_2018} of the diffuse emission in near the Galactic center between $\sim2$~TeV and $\sim20$~TeV, including these secondary contributions from first time.
Here, we improve the modeling of the secondary emission by using state-of-the-art CR propagation codes and adding backgrounds on top of the DM signal.
First, we make use of {\tt HDMSpectra} to obtain the $e^+e^-$ injection spectra. We introduce these spectra into the {\tt DRAGON2}~\footnote{\url{https://github.com/cosmicrays}}~\cite{DRAGON} and run the propagation of the injected particles using the same propagation setup as in the background model (see Ref.~\cite{Luque_PeV} for more details). Finally, we compute the resulting $\gamma$-ray emission with {\tt HERMES}~\cite{HermesCode}. An example of the final $\gamma$-ray emission resulting from a DM decay via the $\tau^+\tau^-$ channel is shown in Fig.~\ref{fig:PromptIC}.

\textbf{Gamma-ray absorption}: 
At high enough energies ($\gtrsim 100$~TeV) the absorption of $\gamma$-ray photons by pair-production ($\gamma \gamma \rightarrow e^+ e^-$) on the CMB and Galactic radiation fields becomes relevant and must be also taken into account. To evaluate the effect of absorption on DM $\gamma$-ray signals, we implemented absorption from the spatially independent CMB photon field in the {\tt HERMES} code, following the same prescription as the one used in the code for diffuse emissions~\cite{HermesCode}. The impact of neglecting the infrared Galactic background, extragalactic background light and star light is very low in those regions of the Galaxy studied in this work, with variations in the total absorption up to only $\sim 10\%$. Furthermore, the description of the spatial distribution and properties of these radiation fields is still quite uncertain and, therefore, we simply neglect them from now on.

\begin{figure}[t!]
\includegraphics[width=\columnwidth]{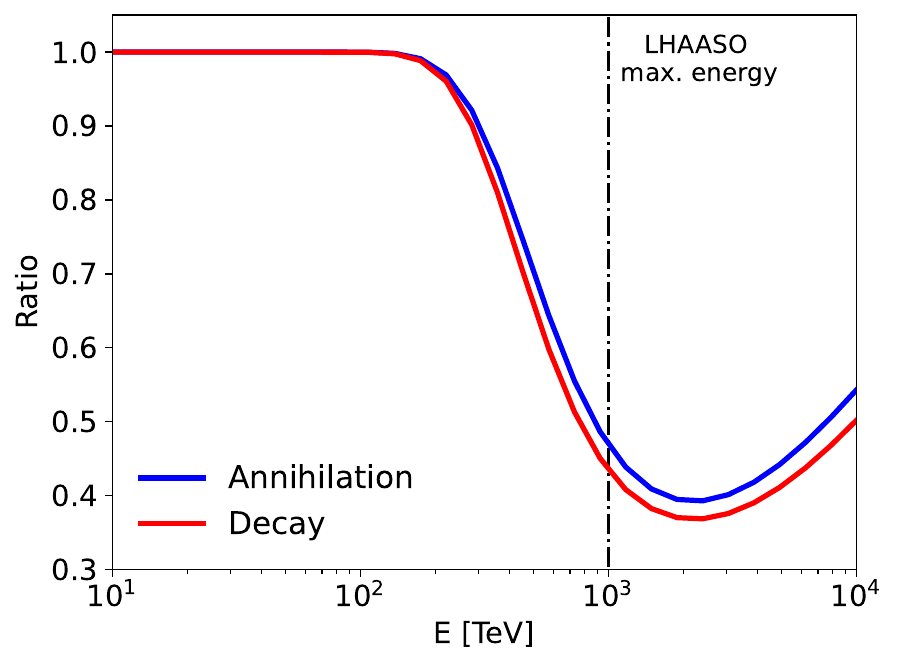}
\caption{Absorption effect in the studied LHAASO region, normalized over the non-absorbed flux. The blue line shows the absorption in the DM annihilation case, whereas the red one shows the absorption for decay. The dash-dotted line shows the maximum energy reached by LHAASO in the dataset we use. }
\label{fig:absorption}
\end{figure}

Given that absorption becomes significant at energies above $\sim100$~TeV (i.e. the maximum energy reported by HAWC), the effect is mostly relevant for the KM2A (LHAASO) dataset. We illustrate this fact for the LHAASO region in Fig.~\ref{fig:absorption} up to $10$~PeV, although we remind the reader that the maximum energy of available LHAASO measurements is $1$~PeV. At this energy, the decrease of the gamma-ray flux can reach $\sim 50\%$ for annihilation and up to $\sim 60\%$ for decay. The maximum absorption is produced around 2 PeV. For that energy, the maximum of the pair-production cross section corresponds to the peak of the CMB photon distribution \cite{Esmaili_2015}. This slight difference in the effect of the absorption between annihilation and decay is due to the fact that, while decay depends linearly on the DM density, annihilation depends on the DM density squared, which highlights once again the importance of an accurately modeling in order to obtain precise and reliable DM constraints.

\textbf{DM limits' procedure:}
Given that the data can be well reproduced by astrophysical background emissions~\cite{DeLaTorreLuque:2025zsv, vecchiotti2024}, in the following we set DM constraints at the $2\sigma$ confidence level, using the chi-square method described in Ref.~\cite{chisquare}. In this approach, the chi-square value $\chi^2$ is calculated as
\begin{equation}
    \chi^2 = \sum_{i \in \lbrace i|\phi^{mod}_i > D_i\rbrace} \frac{\left(D_i-\phi^{mod}_i(x)\right)^2}{\sigma^2_i},
\end{equation}
where $D_i$ represents the data value at the energy bin $i$, $\sigma_i$ its corresponding uncertainty and $\phi^{mod}_i$ is our model prediction, given by
\begin{equation}
    \phi^{mod}_i (x) = \phi^{prompt}_i (x) + \phi^{IC}_i (x) + \phi^{bkg}_i,
\end{equation} 
where $x$ is either the annihilating cross-section or the decay lifetime. The limit is found by varying the cross-section/mean lifetime, and this process is repeated for every annihilation/decay channel and DM mass considered. In this approach, the 2-$\sigma$ limit is obtained when $\chi^2$ = 4~\cite{Cirelli_2023, Delatorre_PBHs}. Only those data at energy bins where the model exceeds the measured flux are considered.

\begin{figure}[bh!]
\includegraphics[width=0.95\columnwidth]{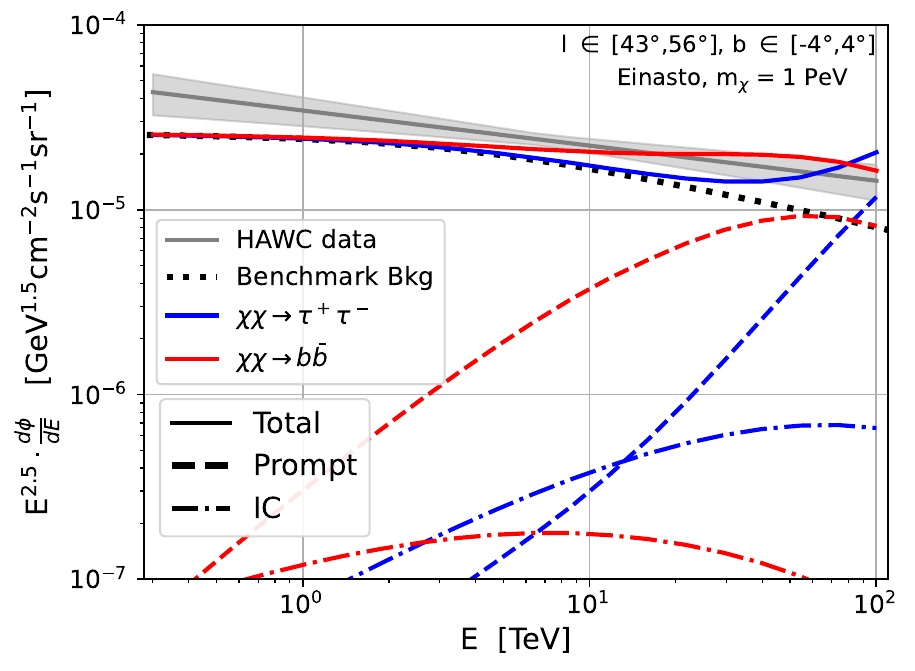}
\includegraphics[width=0.95\columnwidth]{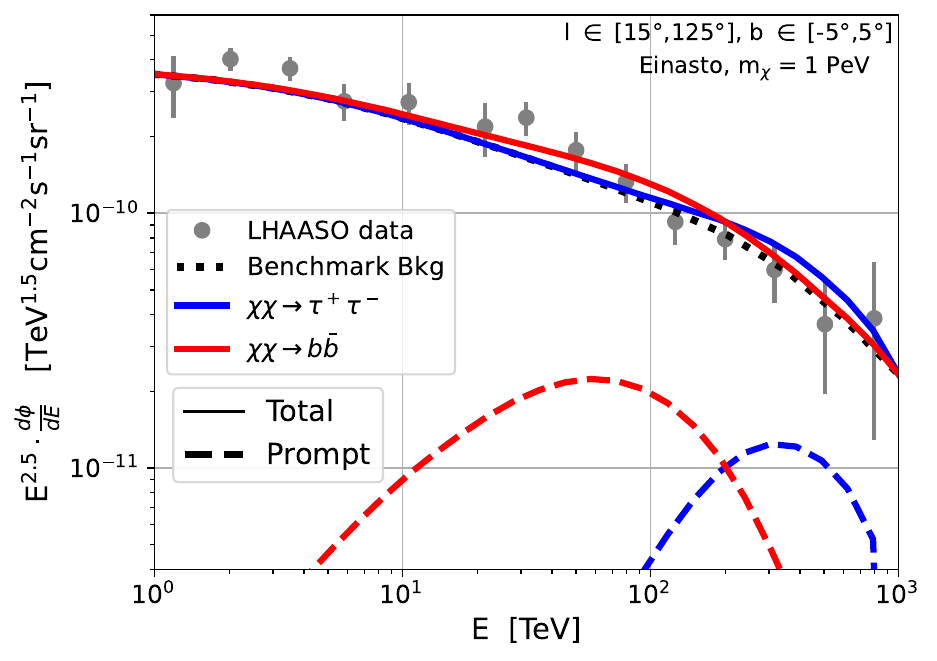}
\vspace{-0.2cm}
\caption{Examples of DM annihilation $\gamma$-ray fluxes for a DM mass of 1 PeV, allowed at the 2$\sigma$ level by the HAWC (top) and LHAASO (bottom) datasets. Blue lines represent annihilation into $\tau^+\tau^-$, while red is for $b\bar{b}$. Dashed lines show the prompt $\gamma$-ray emission, dash-dotted lines represent the IC contribution, the black dotted line shows the benchmark background model and solid lines show the total $\gamma$-ray flux. We do not show the IC component separately for the LHAASO region because its contribution is negligible.} \vspace{-3mm}
\label{fig:fluxes}
\end{figure}

A comparison of the predicted emissions, for the $\tau^+\tau^-$ (blue) and $b\bar{b}$ (red), channels is shown in Fig.~\ref{fig:fluxes}, with HAWC (top) and LHAASO (bottom) data. The dashed lines represent the prompt $\gamma$-ray emission, the dash-dotted lines show the DM-induced secondary IC (only shown for HAWC data since it is negligible for LHAASO), dotted lines for the astrophysical background, while the solid lines are the sum of all these components.

\section{DM bounds from HAWC and LHAASO diffuse observations}
\label{sec:results}

It can be noted in Figure~\ref{fig:fluxes} that both data sets show a power-law trend, in contrast, a DM signal would feature a peaked structure close to the mass of the DM particle, as shown in Figure~\ref{fig:PromptIC}. This kind of smooth power-law behavior is expected from standard astrophysical mechanisms of gamma-ray emission at such high-energies, as already remarked in Refs.~\cite{DeLaTorreLuque:2025zsv, vecchiotti2024}.

\begin{figure}[h!]
    \centering
    \includegraphics[width=0.95\columnwidth]{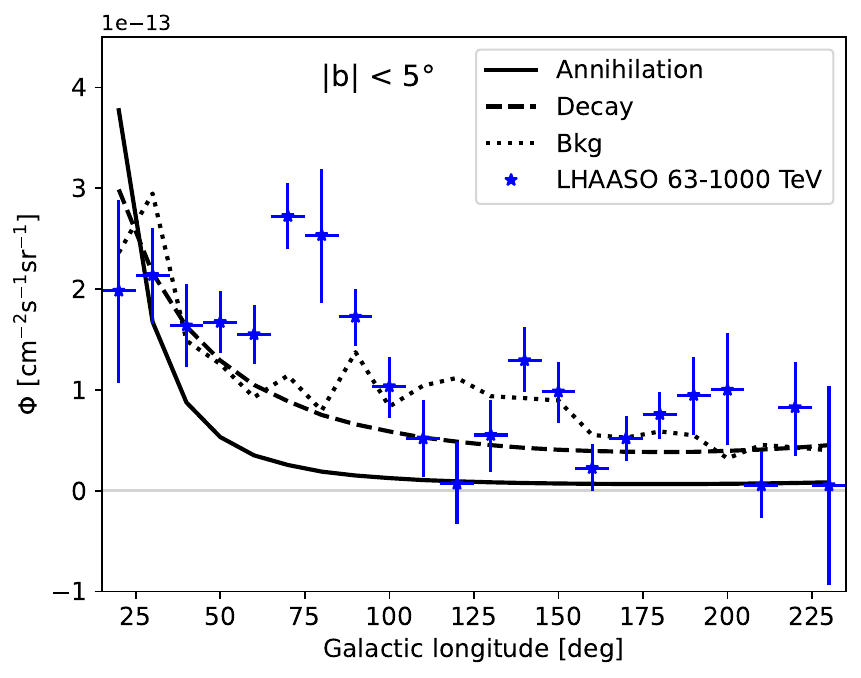}
    \includegraphics[width=0.95\columnwidth]{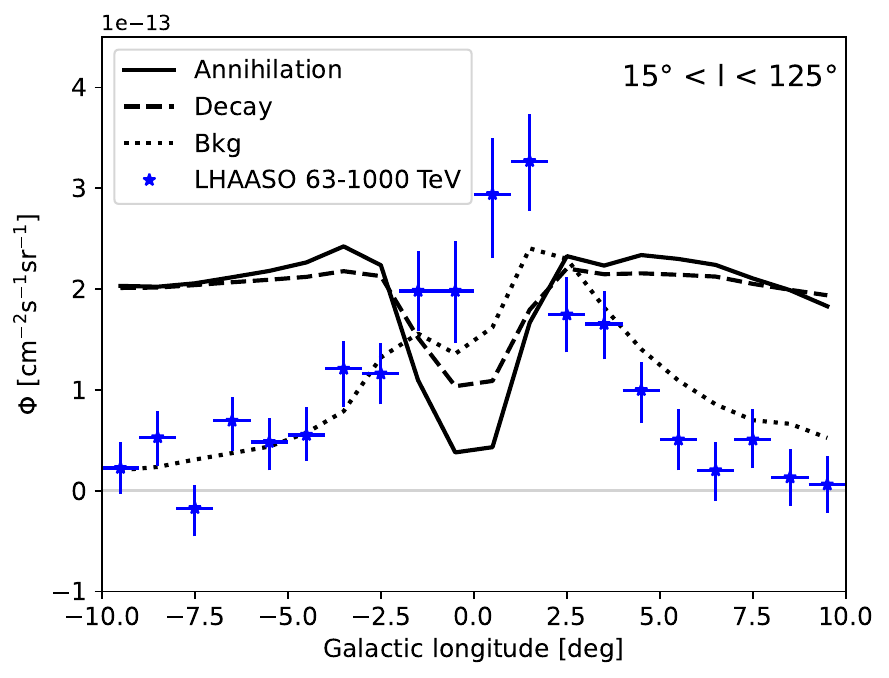}
    \caption{Longitude (top) and latitude (bottom) profiles of LHAASO measurements (blue data points), DM annihilation model (solid line), DM decay model (dashed line) and our benchmark background model (dotted line).}
    \label{fig:LONLAT}
\end{figure}

The first paper that explored DM as the origin of the first PeV $\gamma$-ray emission ever detected was Ref.~\cite{Esmaili_2021}, employing TIBET-AS$\gamma$~\cite{TibetASgamma:2021tpz} data. In this work, they found that the spatial distribution of the events measured by TIBET was incompatible with a DM origin. A similar comparison can be performed with the spatial distribution of the diffuse fluxes reported by LHAASO and HAWC.
Fig.~\ref{fig:LONLAT} shows a comparison of the measured longitude and latitude profiles measured by LHAASO with the expected morphology arising from DM. The upper panel in Figure~\ref{fig:LONLAT} shows the longitude profile extracted from the region with $|b| < 5^\circ$. From this figure, we see that while the expected profile from DM annihilation does not seem compatible with the measurements, the morphology of the emission from DM decay can fit the data well enough and be relatively similar to what is expected to the hadronic emission from CRs. However, this comparison starts at a $l$ = 15$^\circ$. If we extrapolate this morphology to the center of the Galaxy, it would lead to a $\gamma$-ray emission that grows much faster than what current observations tend to indicate. This makes the decay scenario an unlikely possibility.
Furthermore, when looking at the latitude profile extracted from the region $15^\circ < l < 125^\circ$ (lower panel in Figure~\ref{fig:LONLAT}), one can clearly see the very different morphologies that data and DM model present. While the data peaks at the center of the Galactic Plane, as expected, the DM models have their minimum at that position. This is mainly due to the LHAASO mask: 
While the DM emission changes very smoothly with latitude, removing a large portion of the galactic plane causes an abrupt decrease at the central latitudes.
Therefore, when using the LHAASO Mask, the DM flux at very low latitudes is very suppressed.

This incompatibility between the data and a pure DM scenario indicates the necessity of a dominant background Galactic $\gamma$-ray model~\cite{DeLaTorreLuque:2025zsv}. Still, a subdominant contribution cannot be discarded, and therefore, we set $2\sigma$ limits on the decay lifetime and annihilation rates that such DM contribution must have to be compatible with these observations.

By means of the chi-square ($\chi^2$) procedure described in Section~\ref{sec:Method}, we obtain constraints on VHDM using the available LHAASO and HAWC measurements up to masses of $10$~PeV, for different annihilation/decay channels.
In particular, we will show the results for the region providing the strongest constraints, which correspond to $l\in [15^\circ,125^\circ], ~b\in[-5^\circ,5^\circ]$ for LHAASO (for the WCDA and KM2A detectors) and $l\in [43^\circ,56^\circ], ~b\in[-4^\circ,4^\circ]$ for HAWC. These limits are similar to those obtained for the other regions for which HAWC \cite{HAWC_data} and LHAASO \cite{Cao_2023} provided data, but stronger by up to a factor of 2 and 5, respectively. The comparison for $\tau^+\tau^-$ annihilation in the different dataset regions is shown in Appendix~\ref{sec:comp_regs}.

\begin{figure*}[th!]
\includegraphics[width=0.99\textwidth]{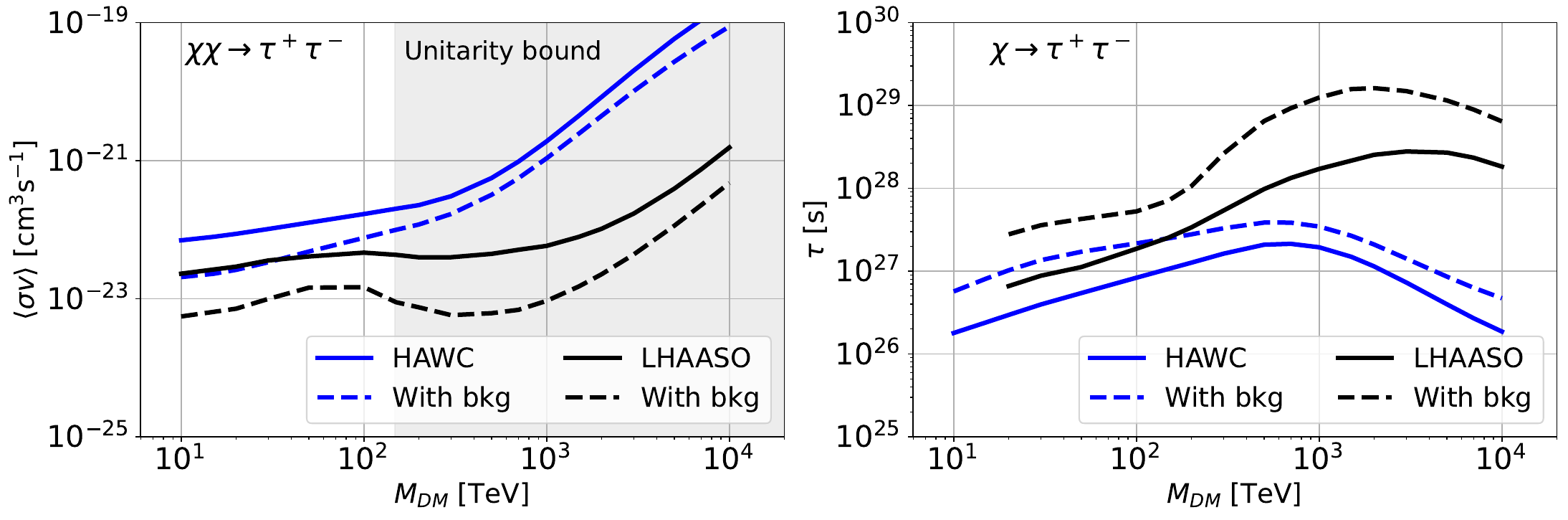}
\caption{DM annihilation cross-section (left) and decay lifetime (right) limits. Blue lines show the limits obtained using HAWC data while black ones refer to those obtained using LHAASO data. The solid lines are our conservative DM limits, while the dashed lines represent our more stringent limits once the astrophysical background model has been taken into account. The shaded area in the annihilation cases represents the region beyond the unitarity bound \cite{Unitarity,Smirnov_2019}. }\vspace{-3mm}
\label{fig:H&L_comp}
\end{figure*}

The 95\% ($2\sigma$) confidence limits for DM annihilation and decay into $\tau^+\tau^-$ are shown in Figure~\ref{fig:H&L_comp}. The solid lines represent our conservative limits obtained using only the prompt emission, whereas dashed lines represent the limits obtained adding the background model.
These limits are obtained using an Einasto profile with $\alpha=$ 0.17 \cite{einasto_pars}, but we also show these constraints, together with constraints in the $b\bar{b}$ channel, for other referent DM distributions. In particular, in Fig.~\ref{fig:prof_uncert} we show how the limits for a NFW profile~\cite{Navarro:1995iw} as well as for other more extreme models of DM distribution: A Moore profile~\cite{Moore_1999}, which predicts a much higher DM distribution around the center of the Galaxy, and a Burkert one~\cite{Burkert:1995yz}, which features a very flat DM distribution in the Galaxy (see Fig.~\ref{fig:dm_profs} and Appendix~\ref{sec:DM_prof} for more details). These comparisons allow us to asses how robust are the constraints derived from these datasets to the uncertainties in DM distribution. As we see, for decay, the uncertainty on the limit for the DM lifetime associated to the DM distribution is negligible ($\sim5\%$), while for the case of annihilation this uncertainty grows to a factor of up to $3$ in the case of LHAASO dataset, and a factor of a few tens of percent in the case of the HAWC dataset. The reason becomes very clear when comparing the different DM profiles in those regions of interest, shown in Fig.~\ref{fig:dm_profs}.

The LHAASO data imposes stronger constraints in the whole mass range for both annihilation and decay. At low masses, LHAASO impose constraints of 2$\cdot$10$^{-23}$ cm$^3$/s, 3 times stronger than HAWC. This factor scales up to 2 orders of magnitude at masses of 10 PeV, where LHAASO impose a limit of around 1.5$\cdot$10$^{-21}$ cm$^3$/s. We also show in Fig.~\ref{fig:regs_lhaaso} a comparison of the limits obtained using the inner and outer regions where LHAASO has reported their measurements of the diffuse $\gamma$-ray emission.

\begin{figure*}[h!]
\includegraphics[width=0.88\textwidth]{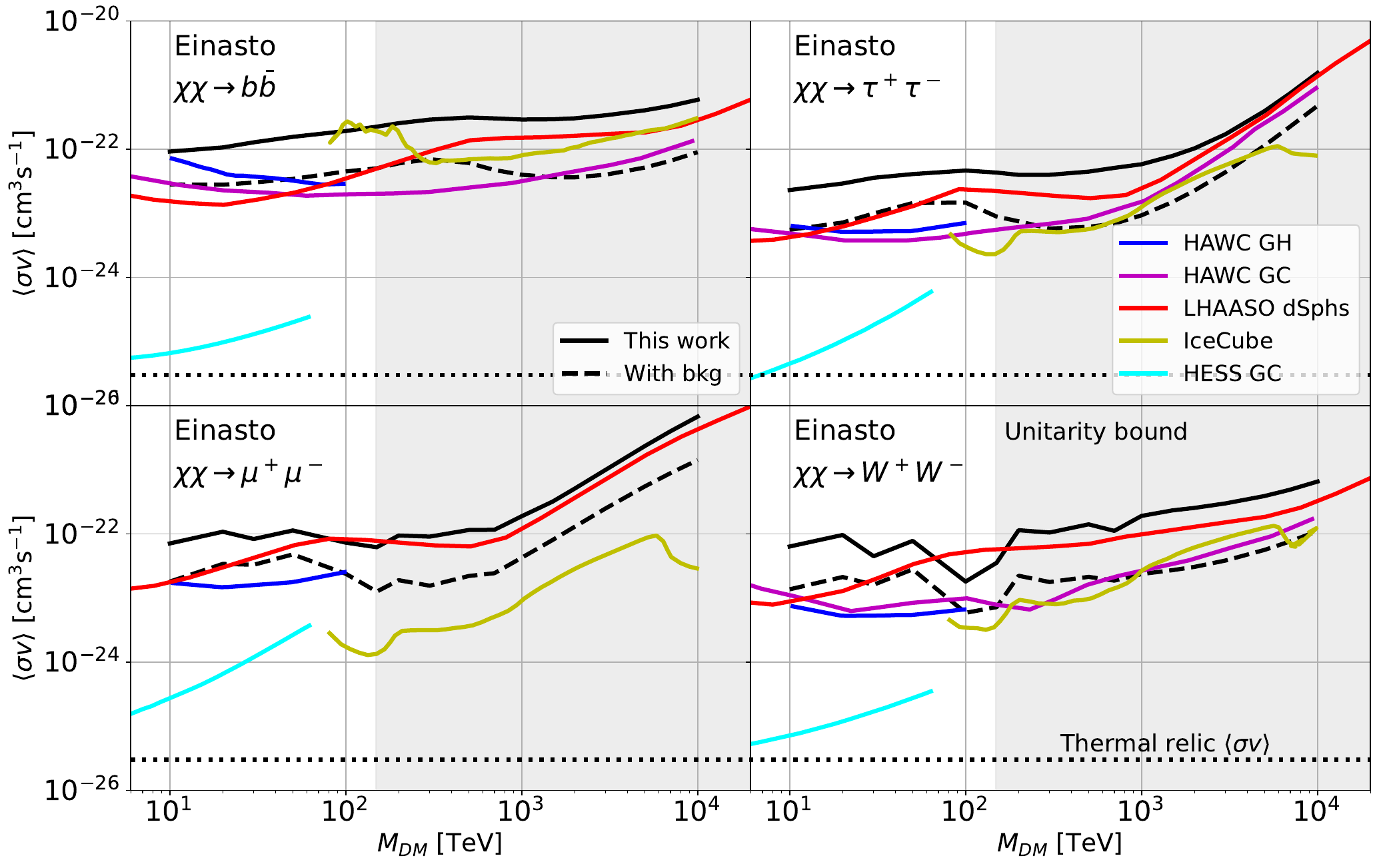}
\caption{Annihilation cross-sections 95\% confidence limits. Black solid (dashed) lines show our conservative (stringent) constraints; blue and purple show the limits by HAWC observations of the Galactic Halo and Galactic Center~\cite{HAWC_GH, newHAWC_GC}, respectively; red shows LHAASO constraints with dwarf spheroidal galaxies~\cite{LHAASO_dSph}; yellow shows IceCube limits using all-sky observations~\cite{IceCube_limits,newHAWC_GC}; and, finally, pink shows HESS limits from the Galactic Center~\cite{HESS_limits}.}\vspace{-3mm}
\label{fig:ann_final}
\end{figure*}

\begin{figure*}[h!]
\includegraphics[width=0.88\textwidth]{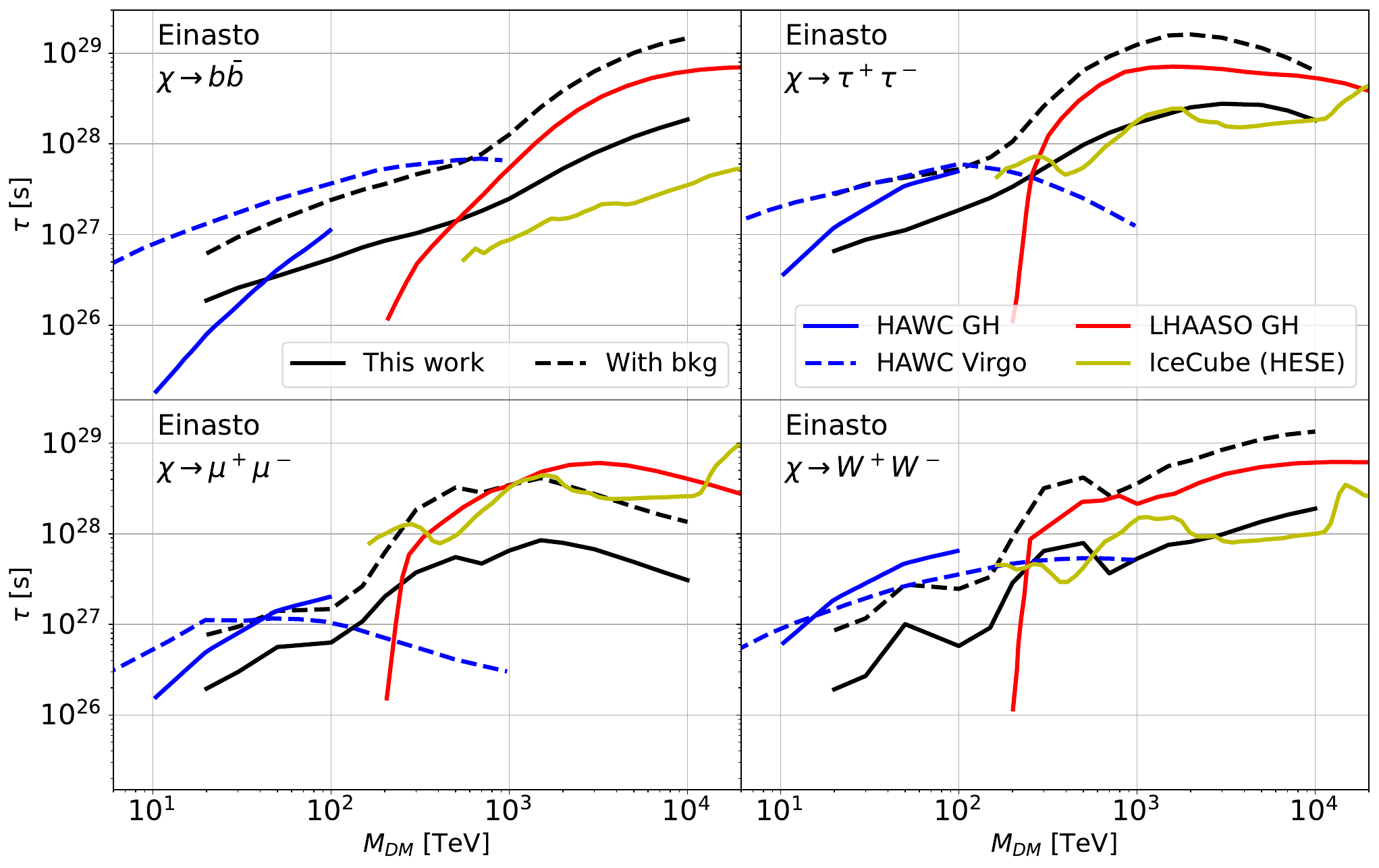}
\caption{Decay lifetime 95\% confidence limits. Black lines show our results, dashed blue shows HAWC constraints with the Virgo Cluster~\cite{HAWC_Virgo}, solid blue shows HAWC limits using observations of the Galactic halo ('GH' in the legend)~\cite{HAWC_GH}, red also shows Galactic halo limits but with LHAASO~\cite{LHAASO_GH}, and yellow shows IceCube constraints using all-sky observations~\cite{IceCube_limits}.}\vspace{-3mm}
\label{fig:dec_final}
\end{figure*}

However, given that it is expected that the diffuse Galactic $\gamma$-ray flux is dominated by interactions of diffuse cosmic rays with the interstellar gas and emission from sources that are not resolved, one can benefit from the use of models of the background emission to set realistic constraints. This not only strengthen the limits obtained for all the channels and in the whole mass range, but it highlights the impact of having reliable background models for DM searches.
The limits including the background model described above are shown with dashed lines. The inclusion of this component allows us to improve the constraints by even an order of magnitude for certain masses, reaching 6$\cdot$10$^{-24}$ cm$^3$/s for the annihilation into $\tau^+\tau^-$ and almost 10$^{29}$ s in both decay channels. As already discussed in Section~\ref{sec:Method}, this background model corresponds to the minimal flux predicted from a fit to lower energy $\gamma$-ray data that only accounts for cosmic-ray interactions, therefore lacking contribution from unresolved sources, and being a conservative choice of background emission that reproduces quite well the trend of the HAWC and LHAASO data.
Using a more aggressive background model can lead larger improvement on our limits for some masses.

Our conservative limits for the annihilation cross section, represented with black solid lines in Figure~\ref{fig:ann_final}, are less constraining than previous limits in the literature, given that the available observations that we are using are located far from the Galactic Center. In turn, when including the Galactic diffuse emission on top of the DM signal, our limits become the most stringent among those from $\gamma$ rays in the high-mass range. This is remarkable given that the regions of the sky that we are using are not optimized to be especially sensitive to DM searches. Moreover, the explored regions also allows for constraints that are less affected by uncertainties in the DM contribution (see Figs.~\ref{fig:dm_profs} and~\ref{fig:prof_uncert}).
We note that neutrino observatories are also highly sensitive to VHDM. IceCube bounds on heavy DM dominate, in general, at medium mass ranges, becoming specially strong in the $\mu^+\mu^-$ channel shown in the lower left panel of Fig.~\ref{fig:ann_final}. However, our limit dominates over 1000 TeV in the $b\bar{b}$ and WW channels. For lower masses, HAWC Galactic constraints are, in general, stronger than ours, and HESS remains much stronger, excluding values of annihilation rate that are unreachable by experiments like HAWC or LHAASO.

When looking at the decay constraints, shown in Figure~\ref{fig:dec_final}, they are also competitive with previous limits in the literature. We provide the most stringent constraints at masses above 100 and 700 TeV for the $\tau^+\tau^-$ and $b\bar{b}$ channels, respectively. Even though HAWC constraints from the Virgo cluster dominate at low masses, this result is notorious given that the regions of the sky we are using were not optimized for DM searches, but for analyses of the diffuse Galactic emission.

\section{Summary and Conclusions}
\label{sec:discussion}

The new measurements of the VHE $\gamma$-ray sky are allowing us to improve our understanding of the processes involving particle acceleration at sources and the interactions of these particles with ambient plasmas and magnetic fields. On top of that, these measurements provide an unprecedented opportunity to probe VHDM models, that may be WIMPs but also particles beyond the WIMP paradigm. Examples for the latter are WIMPIzillas~\cite{Chung_1998, kolb1998wimpzillas, Kolb_2017}, composite DM~\cite{CompositeDM}, glueballs~\cite{Carenza_2023}, branons~\cite{PhysRevD.69.043509, Cembranos_2003}, gravitinos~\cite{Faraggi_2002, Boddy_2014}, among many others~\cite{Hambye_2020, Chianese_2017, Higaki_2014, Contino_2019, Kim_2019}.

Probing this TeV-PeV mass range presents significant challenges for direct detection experiments, which lose sensitivity at such high masses, and for colliders, which struggle to reach the required energy scales. Consequently, indirect detection methods have become crucial in the search for WIMPs \cite{2019_CTA_Book, cirelli2024darkmatter} and VHDM particles, especially via the study of VHE $\gamma$-ray emission. Neutrino observatories are also highly sensitive to VHDM. Stringent bounds on heavy decaying DM have been achieved with IceCube excluding lifetimes of up to 10$^{28}$ s depending on the decay mode \cite{IceCube_limits}.

In this work, we make use of the most recent measurements of the VHE Galactic $\gamma$-ray diffuse emission from HAWC and LHAASO. We present comprehensive calculations of prompt and secondary $\gamma$-ray emissions associated to DM in various annihilation/decay channels. Furthermore, we remark the importance of absorption effects in this VHE range even though we deal with Galactic diffuse emissions, and show how they vary for the cases of decaying or annihilating DM.

The upcoming CTAO will certainly improve these limits, hopefully reaching better sensitivities to DM than the HESS experiment, and extending it up to the PeV mass range~\cite{CTAO:2024wvb}. Given its expected spatial and energy resolution, CTAO will be able to probe the diffuse Galactic emissions with a factor $\sim$10 better sensitivity, offering a powerful tool to probe VHDM beyond any of the current instruments~\cite{CTAConsortium:2023tdz}. The SWGO experiment, which will be similar in concept to HAWC, is expected to probe DM masses and improve the current HAWC constraints by more than a factor 100~\cite{Viana_2019}, reaching masses above the PeV. Especially their searches of the Galactic center will test a significant portion of the DM parameter space and spatial distribution, and will even provide valuable information of the dynamics at the Galactic center.

In conclusion, our study shows the potential of observations of the diffuse $\gamma$-ray emission in the Galaxy to constrain VHDM. We set constraints that are competitive with the strongest neutrino bounds without any optimization of the observation region. In addition, as we show in Appendix~\ref{sec:DM_prof}, these high-longitude regions are not too sensitive to the precise DM density profile that we adopt, which adds an additional layer of robustness to our results when comparing them to limits extracted from the Galactic center, which depends strongly on the assumed DM profile.
The improvement of the limits when including background diffuse emission also illustrates that better estimations of this component could help us strengthen our limits more than a factor of a few.

Just before the submission of this paper, Refs.~\cite{Boehm_2025,Dubey_2025} appeared following a similar approach to the one followed in this work. The calculations in Ref.~\cite{Boehm_2025} include a boosting factor for the DM density profile and an extragalactic contribution, which seems to be subdominant. Ref.~\cite{Boehm_2025} uses the same background model as us, while Ref.~\cite{Dubey_2025} uses a background model obtained directly fitted from the data. Overall, we find that our results are in agreement with their results.

\section*{Acknowledgements}
P.D.L. is supported by the Juan de la Cierva JDC2022-048916-I grant, funded by MCIU/AEI/10.13039/501100011033 European Union ``NextGenerationEU"/PRTR. The work of P.D.L. and M.A.S.C. is also supported by the grants PID2021-125331NB-I00 and CEX2020-001007-S, which are both funded by ``ERDF A way of making Europe'' and by MCIN/AEI/10.13039/501100011033. P.D.L. and M.A.S.C. also acknowledge the MultiDark Network, ref. RED2022-134411-T.

\clearpage
\newpage
\onecolumngrid

\appendix

\section{DM distribution uncertainty}
\label{sec:DM_prof}

\begin{figure}[h!]
    \centering
    \includegraphics[width=0.5\hsize]{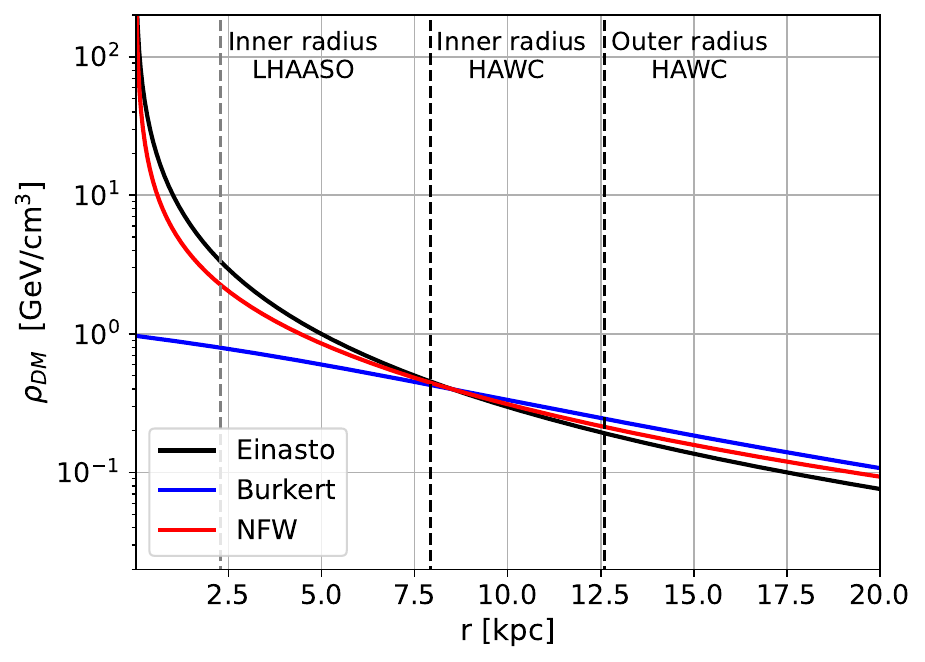}
    \caption{Dark matter density profiles explored in this work. Black shows the Einasto profile~\cite{einasto_prof}, blue shows Burkert~\cite{Burkert:1995yz} and red shows NFW~\cite{Navarro:1995iw}. They are all normalized to a local DM density of $\rho_\odot$ = 0.4 GeV/cm$^3$.}
    \label{fig:dm_profs}
\end{figure}

\begin{figure*}[h!]
\includegraphics[width=0.95\hsize]{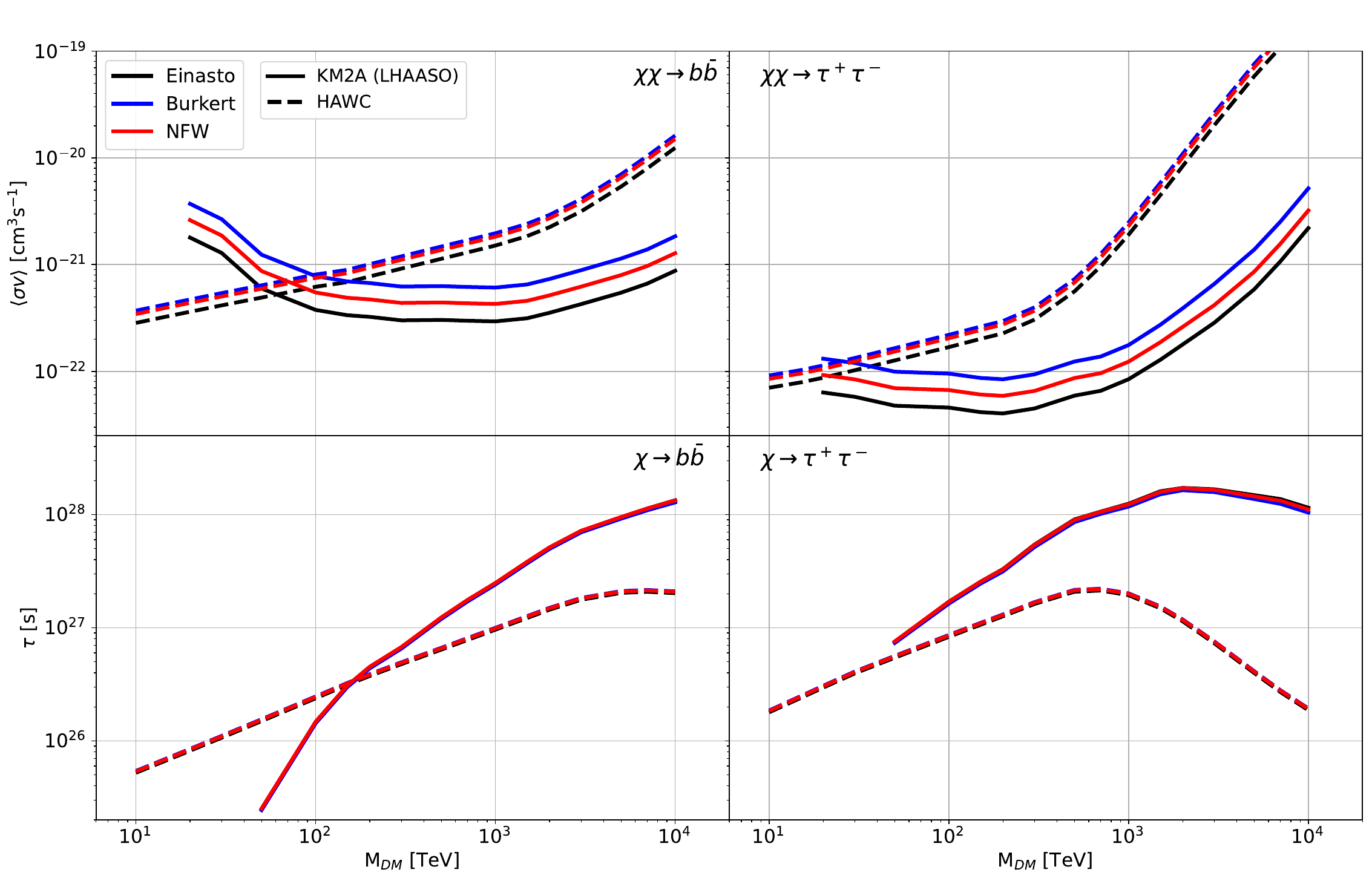}
\vspace{-0.4cm}
\caption{Annihilation cross-section (top) and mean decay lifetime (bottom) 95\% confidence limits for different DM distributions: Black shows our benchmark Einasto profile, blue shows Burkert and red NFW. Dashed lines represent the limits obtained using HAWC data, while solid lines using KM2A data.}\vspace{-3mm}
\label{fig:prof_uncert}
\end{figure*}

A good reason for studying the Galactic Diffuse emission at medium-high latitudes such the ones studied in this work is that the results do not depend strongly on the adopted DM distribution. To illustrate the effect of this uncertainty, we have calculated our limits with different DM distributions: NFW~\cite{Navarro:1995iw} 
and Burkert~\cite{Burkert:1995yz}. They are described by
\begin{equation}
    \rho_{NFW} (r) = \frac{\rho_{s,NFW}}{\frac{r}{r_{s,NFW}}\left(1+\frac{r}{r_{s,NFW}}\right)^2},
\end{equation}
\begin{equation}
    \rho_{Burkert} = \frac{\rho_{s,B}}{\left(1+\frac{r}{r_{s,B}}\right)\left[1+\left(\frac{r}{r_{s,B}}\right)^2\right]},
\end{equation}
where we use 
$r_{s,NFW}$ = 24.42 kpc, $r_{s,B}$ = 12.67 kpc \cite{PPPC4_ID} and both distributions are normalized to the local DM density $\rho_\odot$ = 0.4 GeV/cm$^3$. These DM profiles are shown in Figure~\ref{fig:dm_profs}, while their resulting limits are in Figure~\ref{fig:prof_uncert}. The solid lines show the limits using KM2A (LHAASO) data and dashed lines show using HAWC data. For HAWC, the choice of DM distribution is irrelevant, since the region covered by the data is far from the Galactic Center, and thus mainly determined by the local DM density. In the case of LHAASO, the region is closer to the Galactic Center, which makes the difference between distributions somewhat larger. However, after accounting for the LHAASO source mask, it makes only a slight difference in the resulting annihilation cross-section limits.

\newpage

\section{Comparison regions LHAASO \& HAWC}
\label{sec:comp_regs}

\begin{figure}[h!]
\includegraphics[width=0.95\hsize]{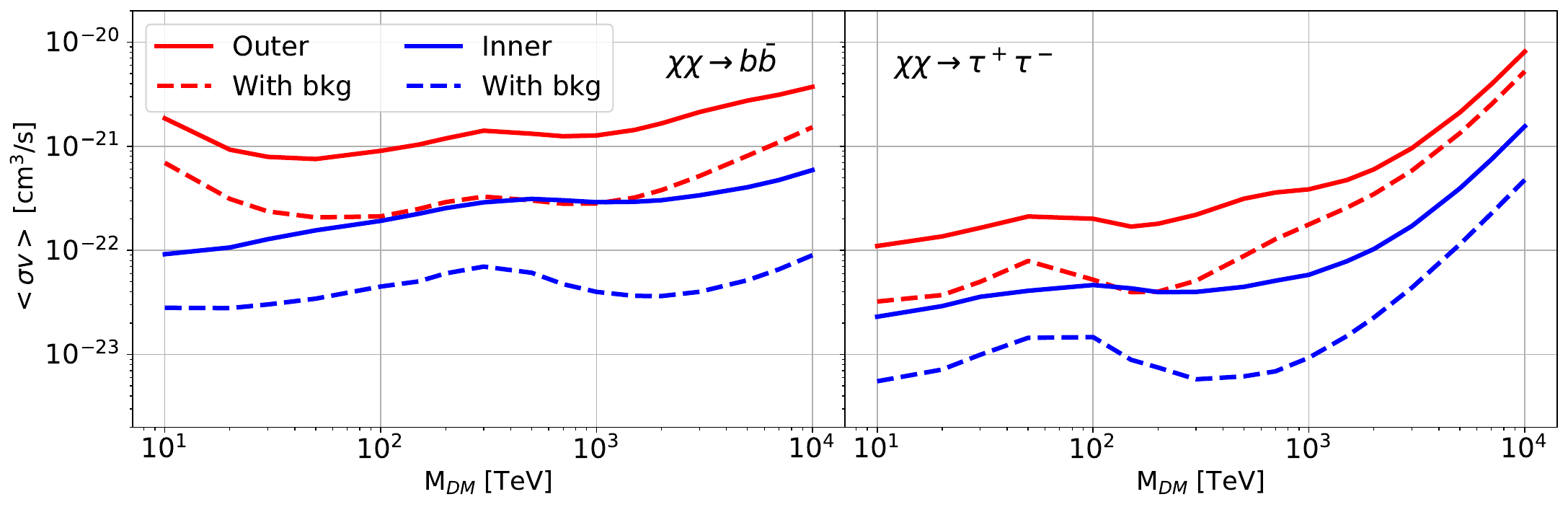}
\vspace{-0.4cm}
\caption{Annihilation cross-section 95\% confidence limits for the different LHAASO regions. Blue represents the inner region (15$^\circ$ $<$ l $<$ 125$^\circ$), while red shows the outer region (125$^\circ$ $<$ l $<$ 235$^\circ$).}\vspace{-3mm}
\label{fig:regs_lhaaso}
\end{figure}

The closer to the Galactic Center, the higher DM density and, therefore, higher $\gamma$-ray production. For this reason, the innermost regions in both experiments give, in general, the most constraining results. This is specially relevant for annihilation because the $\gamma$-ray production depends on the DM density squared, and the innermost region of both LHAASO and HAWC always dominates. We show an example of this in Figure~\ref{fig:regs_lhaaso}, where the limits of both inner and outer LHAASO regions are shown in blue and red, respectively, for comparison.

\begin{figure}[h!]
\includegraphics[width=0.95\hsize]{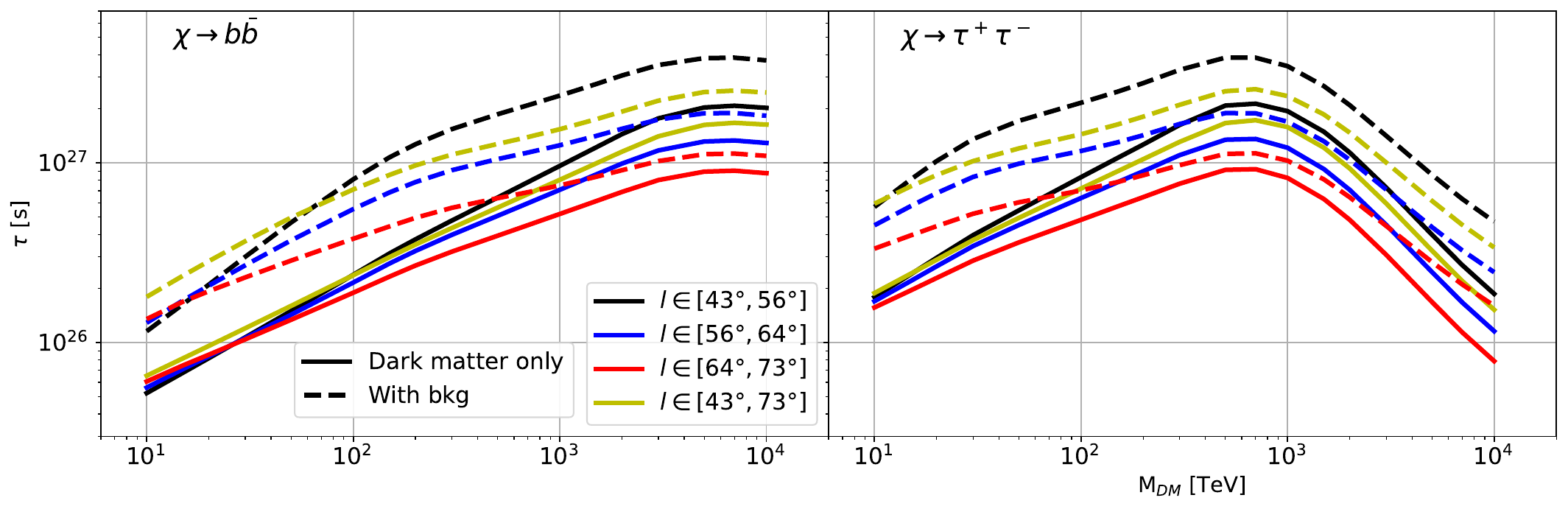}
\vspace{-0.4cm}
\caption{Same as Figure~\ref{fig:regs_lhaaso}, but for HAWC regions.}\vspace{-3mm}
\label{fig:regs_hawc}
\end{figure}

On the other hand, the previous reason is weaker for DM decay, whose $\gamma$-ray production depends linearly on its density. Thus, there are certain cases where the limits could be slightly improved. In the left panel of Figure~\ref{fig:regs_hawc} we compare the limits between the different HAWC regions. The constraint of the whole HAWC region (yellow) can improve that of the innermost one by almost a factor of 2 at the lowest masses considered.

\clearpage
\twocolumngrid
\bibliography{biblio}

\end{document}